\documentclass[11pt]{article}
\usepackage{geometry}                % See geometry.pdf to learn the layout options. There are lots.
\usepackage{graphicx}
\usepackage{amssymb}
\usepackage{amsmath}
\usepackage{fixmath}
\usepackage{MnSymbol}
\usepackage{amsfonts}
\usepackage{array}
\usepackage{multirow}
\DeclareMathAlphabet{\mathpzc}{OT1}{pzc}{m}{it}

\usepackage{caption}
\usepackage{subcaption}

\title{Corrected Wavemaker Theory, Momentum Flux and Vorticity}
\author{Clifford Chafin\\\!~\small{Department of Physics, North Carolina State University, Raleigh, NC 27695} \thanks{cechafin@ncsu.edu}}
\begin{document}
\maketitle
\begin{abstract}
The usual wavemaker theory hides an unjustified integration constant assumption that renders it invalid.  We discuss some surprising subtleties of momentum flux in infinite waves, packets and some counterintuitive examples where ``hidden'' flux contributions can arise.  From here, we construct an ideal wavemaker that builds Airy waves from a series of microjets that allows an exact calculation of the forces on boundaries.  This allows us to also compute the fluxes of all conserved quantities directly and rule out some of these ``hidden'' contributions to momentum flux.  We then discuss some realistic flapper wavemakers and the trouble vorticity sources can create in the resulting fluid motion.  Along the way we present a completely local derivation of the group velocity using fluxes.  
\end{abstract}

The momentum flux is often treated as a fundamental notion in continuum mechanics.  It is used in discussions of elasticity, hydrodynamics and rheology.  Indeed, the Navier-Stokes equations are often considered as momentum conservation equations for the fluid.  In the incompressible, inviscid and athermal case, momentum conservation is sufficient to give a unique set of equations of motion.  Sometimes conservation laws are so restrictive.  For example, Weyl \cite{Weyl} began the effort to show that the stress energy conservation laws in general relativity, $\nabla_{\mu}T^{\mu\nu}=0$ are sufficient to determine that particles follow geodesic motion (up to radiation reaction corrections).  

The Reynold's transport theorem follows from the Navier-Stokes equations as an integral equation \cite{Batchelor}.  Momentum flux is typically extracted from this by associating arguments in the integrals to be the same.  A natural objection would be that some integration constant may have been missed and that this could be some kind of pseudomomentum not suitable for deriving real forces in the medium or on boundaries.  Since this is used in standard wavemaker theory, it is rather important.  

Wavemakers are devices used to mechanically produce waves in the lab.  There has been a long and frustrating history of trying to get theory to match with lab results and getting these in turn to match what is seen in the open ocean.  Even today the topic is a subject of some consternation and some wonder if the theory is missing something important and if the current models of wind driven waves holds some missing features.  

In this article we will examine the notion of momentum flux in detail through particular examples.  Unlike mass, momentum is not a purely advected quantity but has sources in the pressure field of the system.  We will also consider a perfect wavemaker, the kind we believe future systems will aspire to and, through careful examination of the forces, power and torque it produces verify that the usual wavemaker result is only partially correct (and the usual derivation should be considered erroneous for other reasons).  Following this, we will examine some realistic wavemakers and examine the decomposition of irrotational and rotational flow to explain why there has been such variability in results.  

\section{Fluxes}\label{Fluxes}
The amount of mass, momentum, etc.\ that are transported in waves play a crucial role in the discussion of the interaction of waves with each other and external bodies.  Unfortunately, the simple intuitive meaning we draw from the stress tensor is not always correct.  Given that inviscid hydrodynamics is driven only by pressure and no stress appears on might wonder to what extent ``radiative stress'' has meaning in wave theory.  It has been used extensively in the theory of wave set-up \cite{LH}.  Reynolds' stress is a term in a rearranged time averaged N-S equation and there is much debate about its use in turbulence.  One might wonder when, taken on its own, it gives a real stress or pseudostress in the system.  To this end, we consider a set of of examples meant to clarify momentum flux and the flux of other conserved quantities.  

We will build our discussion for the case of nonrelativistic media where the energy, momentum\ldots are entirely in the motion of massive particles of fixed mass and number.  Examples involving contributions from radiation, as in EM fields in a dielectric, are not considered because radiation in media is a special intrinsically relativistic case.\footnote{For a discussion of this see Chafin \cite{Chafin-em}}.  Potential forces are allowed and other forces (such as pressure) that can store energy or transmit force elastically.  This lets us rely on the following (obvious?) principles to check and derive results.  

\begin{itemize}
\item  All momentum in the system is in the sum of the momenta of the constituent particles.
\item  Momentum flux of a compact packet is just the momentum of the packet times the velocity of its motion.  
\end{itemize}

First let us consider the simplest case, mass flux $\mathit{f}_{m}$.  The mass crossing a surface per unit time is the surface integral of the mass density times the velocity:
\begin{align*}
\int_{\text{A}}\mathit{f}_{m}=\int_{\text{A}}\rho v ~da=\int_{\text{A}}\mathpzc{p} ~dz dy
\end{align*}
which immediately establishes the relationship between mass flux and momentum density.

This example was quite easy because both mass is a locally conserved quantity; mass is a completely advected quantity.   The mass motion is really what defined the fluid velocity field in hydrodynamics, so this is true by definition.  Some quantities are almost entirely advected in a moving fluid.  Entropy, temperature, concentration of solutes can diffuse but this is generally quite slow compared to being dragged with the flow.  Unfortunately, the other conserved quantities we care about, e.g.\ energy and momentum, are often not advected but transported in a nonlocal fashion by the pressure field or, as we shall see, unusual actions at the boundaries.  
%This field acts as sources and sinks for these fluxes making the task of constructing a local definition of momentum flux rather subtle.  %(We shall see in Sec.~\ref{Stress} that it is generally not a well-defined proposition.)  

Now let us consider the case of momentum flux.  
The usual definition of momentum flux density is $\Pi_{ij}=p\delta_{ij}+\rho v_{i}v_{j}$ \cite{LL} which is chosen so that 
\begin{align*}
\frac{\partial}{\partial t}(\rho v_{i})=-\frac{\partial}{\partial x_{j}}\Pi_{ij}
\end{align*}
This definition ensures momentum conservation but it is not really clear that it is the rate of momentum density of the fluid passing through the system.  It is ambiguous up to a (spatial) constant and a curl, 
$\epsilon_{jlm}\partial_{l}J_{im}(x,z,t)+u_{ij}(t)$, 
where both the space and time dependence of this field is completely arbitrary and even a gradient of harmonic functions if the boundaries allow it to be nonzero, 
$\partial_{j}\Psi_{i}(x,z,t)$.  
While we can get internal consistency with this definition we can't be sure we don't actually have a quasimomentum - something that is not scaled properly to combine with momentum sources outside the fluid.\footnote{An example of this where we are simply lucky, is the derivation of Poynting's theorem from the free field Maxwell's equations \cite{Jackson}.  In this case, we ignore the overall scale factor and arbitrary curl and come up with the right momentum density to compare correctly to the momentum of the external matter fields.  A lagrangian approach involving EM and matter fields shows that this is the correct answer.}  

We can find the true local flux for an incompressible fluid by adding up the momenta of the constituent particles of the fluid.  %There is no pressure induced contribution to the momentum so there is no reason to consider it part of the flux.  
\begin{align*}
\int_{\text{A}}\mathit{f}_{p}=\int_{\text{A}}\mathpzc{p} v da
\end{align*}
where we take the absolute value of $\mathpzc{p}$ because the flux must be going in the direction of the vector $v$.\footnote{This is based on the (hard to dispute!) assumption that mass is always positive.}  
We can make the connection to the above stress energy tensor by noting that $\mathit{f}_{p}^{i}=(\rho v^{i}v^{j})\hat{v}_{j}$.  This is the part of the stress tensor that does not involve the pressure projected along the direction of fluid motion.  
This suggests we rewrite the conservation law for momentum flux density as 
\begin{align}
\frac{\partial}{\partial t}(\rho v_{i})=-\frac{\partial}{\partial x_{j}}(\rho v_{i}v_{j})    -\frac{\partial}{\partial x_{i}}P
\end{align}
which follows from the N-S equations.  
For pressure functions that vanish at infinity this still gives global conservation of momentum using this new definition of flux.  For a more general procedure that uniquely partitions advected from nonadvected transport of a flux, see App.~\ref{Flux}

\begin{figure}[!ht]
   \centering
   \includegraphics[width=3in,trim=0mm 100mm 0mm 60mm,clip]{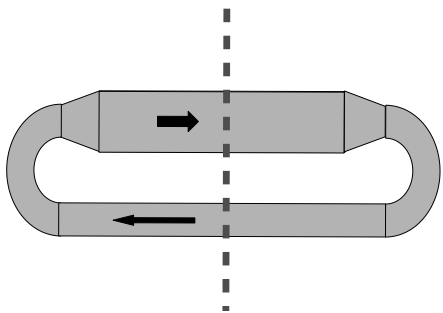} 
   \caption{Flow of an incompressible liquid through a loop with pipes of two diameters.}
   \label{tubes}
\end{figure}

Lets see how this works with a hydrodynamic example.   
Consider loop of two tubes of different area but same length, as in Fig~\ref{tubes}.  Assume steady liquid flow and that the larger tube is twice the area of the smaller one.  Now let us use this definition to calculate the flux through the area that slices both tubes.  
Clearly the net momentum of the system is zero.  The smaller tube carries half the volume of flow but at twice the velocity.  Computing our momentum flux density across the dotted boundary gives: 
\begin{align*}
\mathit{f}_{p}=\int_{\text{2A}}\rho v^{2} da-\int_{\text{A}}\rho (2v)^{2} da=-2\text{A}\rho v^{2}
\end{align*}

This is clearly not zero despite that our steady state system has no net momentum.  What has gone wrong?  The ends both experience an outwards force to redirect the flow.  These balance and contribute to tension in the tube so these are not the solution.  At the flanged regions the narrow flow expands and gets redirected forwards.  This gives a backwards impulse on the left flange.  Similarly, we get a forwards impulse on the right one.  This establishes a tension 
across the tube and creates a higher pressure region in the fluid.\footnote{The case of a gas is special in that there is no microscopically elastic contribution so one could, in principle, track the momentum via collisions.  In a liquid, there are elastic forces transferred over long distances by bonds so no such microscopic counting is helpful.}  The flange regions then create sources and sinks of momentum flux in the tube and the fluid.  It also shows that we shouldn't expect momentum flux conservation in the fluid alone.  
An incompressible fluid or solid can transfer forces at infinite speed.  

It is frustrating in trying to keep track of the fluxes other than the mass \textit{density flux} (=momentum density) of waves because there is no completely local way to describe evolution of the flux.  Mass is never locally destroyed or transported in such an abrupt nonlocal manner\footnote{The mass flow is what actually defines the velocity field that we use to define advection.} but linear and angular momentum can be.  Furthermore, unlike mass, there is no way to tag a parcel of momentum as it moves through the fluid.  Incompressibility does not introduce any ``acausal'' transport of mass but does for these other quantities.  We don't have an analogous problem with mass density flux because the pressure cannot contribute anything to it or transport it in any way but advection.   

\begin{figure}[!ht]
   \centering
   \includegraphics[width=3in,trim=0mm 100mm 0mm 50mm,clip]{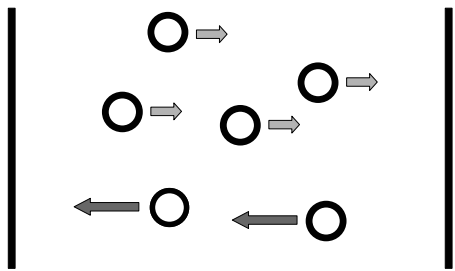} 
   \caption{An anisotropic 1-D gas distribution driven by forces at the walls.}
   \label{gas}
\end{figure}

Let us consider an even simpler example.  A ``gas'' of noninteracting particles exists between two plates, separated by length $l$, as in Fig~\ref{gas}.  The motion is constrained to be 1-D with the following anisotropic velocity distribution.  The particles moving left move twice as fast as those moving right and there are half as many.  This is like the previous situation without any complication of nonlocally determined pressures.  The walls are active in that they collide with each particle with an impulse so that the velocity distribution is a steady state.  
This gives a net particle density of $n=\frac{3}{2}\frac{n_{R}}{l}$ and right momentum density $\mathpzc{p}^{R}=m n_{R}v_{R}=-\mathpzc{p}^{L}$ where $v_{R}$ and $n_{R}$ are the velocities and number of right movers respectively.  
Therefore the momentum density is zero and the momentum flux density is $\mathit{f}_{p}=\mathit{f}_{p}^{R}+\mathit{f}_{p}^{L}=3mv_{R}^{2}n_{R}/l$.
%-mv_{R}^{2}n_{R}/l$.    
The collision rate at each wall is $\nu=v_{R}/(l/n_{R})$ so the force (``pressure'') on each wall is inwards and has magnitude $F=3 m v_{R} \nu=3 m n_{R} v_{R}^{2}/l$.%\footnote{The net force on the system is $2F$ but the relation $M_{net}\ddot{X}_{cm}=2F$ is clearly false since the center-of-mass is not moving.}

%THE ELEVATION OF A PACKET MAY CONTRIBUTE TO THE MOMENTUM FLUX.  

%THE ENERGY SOURCE AT THE WALLS GIVES A HIDDEN FLUX ALSO A HIDDEN MOMENTUM FLUX SOURCE.  THIS GIVES A ZERO MOMENTUM DENSITY WITH A NONZERO FLUX.  THIS IS INCONSISTENT WITH A GROUP VELOCITY.  A PACKET GIVES A HIDDEN END CONTRIBUTION.    

Each collision on the right is imparting $\frac{3}{2}mv_{R}^{2}$ of KE and each on the left subtracts the same amount.    This gives a KE flux of $\mathit{f}_{KE}=-\frac{3}{2}mv_{R}^{3}n_{R}/l$ across the cavity which matches the rate of sourcing of KE at the right wall.  The KE density of the cavity is $\frac{3}{2}mv_{R}^{2}n_{R}/l$.  This gives a relation between the momentum flux and energy density $\mathit{f}_{p}=2\mathcal{E}$.
If we consider the case of an isotropic gas we see that the net momentum density is zero.  (In the RF we always have $\frac{\mathpzc{p}_{L}}{\mathpzc{p}_{R}}=1$).  If we label the KE of right an left moving particles as $\mathcal{E}_{R}$ and $\mathcal{E}_{L}$ we can see from the relation $\mathcal{E}_{R}=\frac{1}{2}\mathpzc{p}_{R}v_{R}$ that the condition for an imbalance in momenta fluxes is that $\frac{\mathcal{E}_{R}}{\mathcal{E}_{L}}\ne 1$.  

%The extra force ($F=3 f_{p}=f_{p}+\Delta F$) at the walls indicate how much extra force must be applied to maintain this.  It does not, however, indicate an extra momentum flux through the gas.  This is analogous to the forces exerted across a rigid body which is elastically compressed to carry force from one side to another.  The walls of the gas are acting as sources and sinks of momentum.    

From here we can see that any particle distribution, in its rest frame, that transports momentum density across it must have an anisotropic distribution of velocities.\footnote{This should make us concerned that our hypothesized notion that momentum flux for packets may not always be the momentum density times the packet velocity.  This turns out not to be the case.}  
If we were to suddenly freeze the plates in place, the system would produce a four velocity distribution function $\{\pm v_{R}, \pm v_{L}\}$ instead of two $\{v_{R}, v_{L}\}$ with net momentum still zero but no momentum flux.  This shows how a sudden change in the behavior at the boundaries can radically alter the forces there and the fluxes \textit{across} (not ``through'') the system.  Consider a solid rod with equal forces $F$ on either end it is compressed and at rest.  Let us consider the same rod under the same pressure but now pushing a large mass.  It is transferring momentum to the body but there is nothing \textit{locally} in the rod that indicates how much.  The forces at the boundary and momentum transferred there is a function of {behavior} of the system at the edges not a locally measurable flux that is transferred through the system.\footnote{There are many cases in physics where a cleverly crafted example hides important contributions at the boundary.  One of the most famous is the ``hidden electromagnetic momentum'' of charged parallel plates in a B-field.  The fringe fields completely cancel the easy to compute internal $E\times B$ contribution on the interior.}    

Let us return to the specific example of water waves.  Since progressive waves have a symmetric distribution of particle velocities, we can be sure the forces exerted at some artificial parcel ends are entirely due to the pressure.  Momentum flux is therefore only the hypothesized bulk drift of momentum due to net mass motion.  
To evaluate the momentum transport for waves, we have to confront the role of pressure and that forces can drive the energy forwards much faster than the particles are moving.  The simplest is wavepacket analysis.  Since we are considering the fluid parcel to be of constant density we see that any net motion of a parcel must be accompanied by a net elevation change \cite{Chafin-rogue}.  

%
%By building packets of compact support we know that there is no hidden transport of momentum through this system in finite or infinite wavetrain case (see Sec. \ref{LH}).
%
%THE ROLE OF ELEVATION OF PACKETS AND DIAG SHOULD GO HERE.  

%From the mass conservation consideration in previous sections, wish to consider elevated packets as in Sec.~\ref{Packets}; specifically we want to have long packets prepared with any necessary elevation changes for mass transport already present so the only changes are spreading and very long wave corrections we can neglect for time scales that are still much longer than the characteristic oscillation time of the waves. 
 
For the case of a packet of amplitude and frequency $(a,\omega)$ and persistently zero outside its support, we find it needs an elevation to transport mass.  The mass that disappears at the back must get circulated back towards the front but regardless of the details of this process we can immediately calculate the net momentum from the elevation and the group velocity.  Matching momentum density with that of a packet elevated by $h$ moving at $v_{g}$ we find $\frac{1}{2}\rho a^{2}\omega=\frac{1}{2}\rho h \omega/k$ so $h=a^{2}k$, where we have assumed a uniform change in elevation.  This implies a mass density change (mass per area) of $\mathit{m}=\rho a^{2}k$ and a mass flux of $\mathit{f}_{m}=\frac{1}{2}\rho a^{2}\omega$ for a packet.

The total momentum in a long packet is unchanged for a long time.  The packet moves at $v_{g}=\frac{1}{2}\sqrt{g/k}$.
From this we can derive the momentum transport rate (for a right moving wave) $\mathit{f}_{p}= \mathpzc{p}v_{g}=\frac{1}{4}\rho g a^{2}$.  This indicates that the the force to create the waves is in the direction of wave advance.  
Angular momentum flux is found similarly:  $\mathit{f}_{L}=\mathcal{L}v_{g}=(-\frac{1}{4}\rho g\frac{a^{2}}{\omega})\frac{1}{2}\sqrt{g/k}=-\frac{1}{8} \rho g\frac{a^{2}}{k}=-\frac{1}{8} \rho a^{2}\frac{\omega^{2}}{k^{2}}$.

The consequence of these relations is that any wavemaker that is designed to make an Airy wave must input force and torque to the wave at these rates.  There are no hidden contributions by the above.  There will be other pressure driven stabilizing contributions as as shown in Sec.~\ref{Local}.  Applying force and torque to the water surface does not guarantee that it will all end up in the wave.  Significantly, one also has to inject a component of fluid according to the above mass flux.  This means that none of the simple flapper tank designs can generate such waves.  Their waves will always involve backwards component waves, backwards circulating flows and long surface elevation gradients (see Sec.~\ref{Wavetanks}).

\section{A Local Derivation of the Group Velocity}\label{Group}
It has long been known that the group velocity is also the velocity of energy transport \cite{Lamb}.  It follows from the dispersion relations for large wavepackets.  How to see this from a local point of view is puzzling.  We will see one way to tackle this in Sec.~\ref{Local} explicitly at the wavemaker.  Here we look for a more geometric approach that is an application of intuition derived from kinetic and potential energy and mass transport imparted per cycle by the ideal wavemaker we discuss Sec.~\ref{Local}.  The advantage of doing this first is that it will give us additional intuition as to the meaning of some some of the terms that arise in calculating the force at the wavemaker.  

In the case of electromagnetic waves in media \cite{Chafin-em}, the plane waves can be thought of as free EM waves that carry all the momentum and oscillating charges that store and release energy.  At the leading edge of packets, the medium picks up some of this momentum and there is a net stress in the medium due to a ``hidden'' standing wave exerting forces on the boundaries and packet ends as well.  This allows nice decomposition of the wave and medium energy and one finds that the group velocity of the packet satisfies the simple relation $\mathpzc{p}=\frac{\mathcal{E}}{c^{2}}v_{g}$.  In surface waves, such a relation is less interesting because of the nonrelativistic nature of the system.  

The wave fronts of an infinite wave advance at $v_{ph}$ while the packets move at the slower $v_{g}$.  When looking at a wave locally there is nothing that indicates the value of $v_{g}$ yet it seems to control the transport of the energy, mass and momentum.  So far, the group velocity is only a function of the rate at which the dispersion relation changes in momentum space, $v_{g}=\frac{\partial \omega}{\partial k}$.  We also know that what is ``local'' in momentum space is very nonlocal in real space.  
It seems that we should be able to locally determine the fluxes of these quantities and so infer a value for $v_{g}$.  This would provide a local derivation for what is generally expressed as a nonlocal quantity.

As a first complication, we know that pressure in an incompressible fluid is able to instantaneously transmit  momentum flux over arbitrary distances.  Given a long tube filled with an incompressible fluid when we push on one end, the force is instantly felt at the other end.  We now understand this as the fluid having sources and sinks of momentum flux at each end.  Nevertheless, this makes us wonder about how locally we should look at the waves for these fluxes.  If we constrain ourselves to periodic waves, then we can see that we won't have such pressure build up and nonlocal transfer of forces over longer than a wavelength.  This suggests that the wavelength is the natural scale to look at for transport of these quantities.  

\begin{figure}[!ht]
   \centering
   \includegraphics[width=3in,trim=0mm 0mm 0mm 0mm,clip]{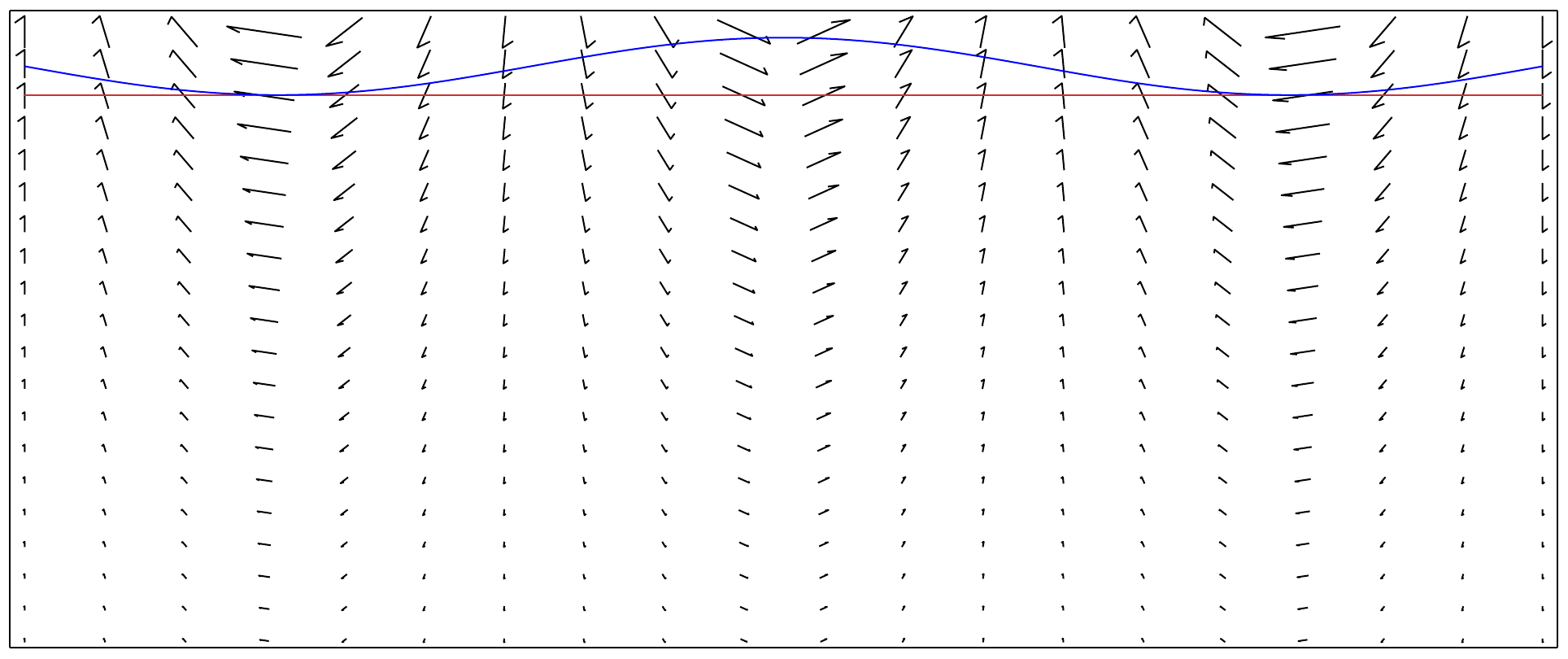} 
   \caption{The velocity profile of a wave superimposed upon its surface elevation profile.
   }
   \label{wavecut}
\end{figure}

For the case of energy transport let us consider kinetic and potential energy separately.  When we look at the kinetic energy density of the wave we find it is nearly uniform.  Peaks have a slight excess $\delta KE=\int_{-a}^{\eta}\frac{1}{2}\rho v^{2}$ but most of this remains below the troughs where it is uniform as in Fig~\ref{wavecut}.  Potential energy, however, is entirely decided by the motion of the crests.  The PE is transported at $v_{ph}$ because it is a locally defined quantity attached to the wave shape.  This means that, to lowest order, the energy flux is 
\begin{align*}
\mathit{f}_{E}=\mathit{f}_{PE}=\mathcal{U} v_{ph}=\frac{1}{4}\rho g a^{2} \sqrt{\frac{g}{k}}
=(\mathcal{E}v_{g})
\end{align*}
We see that the potential energy is transported by the crests but the kinetic energy is essentially uniform and transported no faster than the Stokes drift.  This means that the potential energy has a flux and the kinetic energy does not.  We conclude that, to leading order, waves transport mass, linear and angular momentum and \textit{potential} energy.

At the backside of a finite packet, the remaining kinetic energy pushes up a smaller wave that shares the net energy more evenly.  Each passing smaller crest at the edge successively removes potential energy until all the kinetic energy is gone as well.  This damps the wave size by half with each passing crest.  An opposite behavior occurs at the leading edge of the packet.  This gives a nice geometric explanation for why the group velocity is half that of the phase  velocity in deep water.  

%For the case of elevated mass of a packet, we know that the wavefronts advance at $v_{ph}$ and are created and vanish at the ends.  If mass advanced at this rate it would pile up at the end of the packet and deplete at the other.  The only consistent choice is for the excess packet mass to move at $v_{g}$.  This means we can assume a uniform mass elevation in the flat interior of the packet and that the wave crests carry half of each away with each pass.  At the tapered back end of the packet, the kinetic energy pushes mass up and forwards, this lowers the mean elevation to the equilibrium height as the crest height vanishes.  

This example lets us see how locally conserved fluxes of waves can determine the group velocity.   
In general, whenever we have conserved quantity $\mathcal{C}$ that has a fraction bound to the moving crests, we can derive the group velocity directly from the phase velocity and the fraction of energy moved by the advancing crests.  
\begin{align}
v_{g}=\frac{\mathcal{C}_{\text{bound}}}{\mathcal{C}_{\text{net}}}v_{ph}
\end{align}
This result is a feature of the conservation laws and wavepacket motion and independent of whether the system is linear or not.  If would be interesting if we could derive a converse to this; specifically to give a way to assign a fraction of any conserved quantity unambiguously to the crest motion.

\section{Ideal Wavemakers}\label{Local}

The kind of flapper wavemakers that include damping surfaces with wave breaking have some limitations.  They invariably introduce vorticity at the flapper and the breaking surface as well as the walls.  Angular momentum is transferred to the ``shore'' in an asymmetrical fashion; first at depth and then in the breaking of the wave and run up of the flow onto the ground.  For fluid motion that contains little net drift but strong oscillatory motion, this can create strong flows and sources of vorticity \cite{Chafin-acoustic}.  Even though such motion is generally far slower and shorter lasting than the waves themselves, they can often propagate over the full scale of a laboratory setup and create confusing results.  Ultimately, we can decompose the flow into irrotational and rotational components by the Helmholtz decomposition but the sources and sinks of energy into this rotational motion is hard to model and will be discussed in Sec.\!~\ref{Wavetanks}.

\begin{figure}[!ht]
   \centering
   \includegraphics[width=3in,trim=0mm 50mm 0mm 30mm,clip]{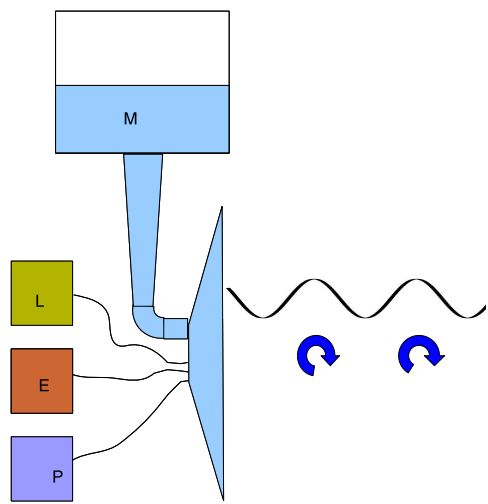} 
   \caption{An ideal wavemaker.  Mass, momentum and energy are imparted through a series of microjets to construct the advancing and rotating motion of particles uncomplicated by backflows and vorticity.  }
   \label{wavemaker}
\end{figure}

Surface shear is known to have a strong effect on wave motion.  
If we want to generate the kinds of waves describes by Airy, Stokes or Gerstner we must have a smarter kind of wavemaker.  Given the fragility of irrotational motion when interacting with solid surfaces, this requires quite a detailed engineering of the local forces and fluxes at the origin of the wave.  In our ideal wavemaker below we describe these local processes as assembling the wave with a series of small jets that inject and withdraw fluid at the vertical line on the left in fig.\!~\ref{wavemaker}.  The great advantage of this method is that it lets us derive all the relevant fluxes directly and instantaneously.  From the relationship between the group velocity, fluxes and densities of quantities carried by wave packets,\footnote{We have carefully worded this because of the Gerstner wave case.  A Gerstner wave packet does not transport  vorticity with it like the other waves transport energy, mass, etc.  In fact, it requires a body of water that contains a backwards shear flow if it is to propagate as a Gerstner packet at all.  See App.~\ref{Gerstner}.} as shown in Sec.~\ref{Local}, we can derive the densities of each quantity from there.  

Consider there to be a reservoir of fluid at rest and at zero potential energy and a progressive wave that grows out of a perpendicular vertical surface.  The particles of fluid that advance or retreat come from or return to the reservoir of fluid.  We provide sources of mass, momentum, angular momentum and energy that this requires at each step as in Fig~\ref{wavemaker}.  The energy to speed a parcel from rest to match the fluid motion is  obvious.  Less obvious is that there is a variation in the pressure field at the surface which is being 
displaced to insert the new fluid parcels.  This will turn out to give a alteration to the standard makemaker analysis of momentum flux.  

We calculate the mass flux per time of an Airy wave by 
\begin{align}
\mathit{f}_{m}&=\frac{\omega}{2\pi}\int_{0}^{2\pi/\omega}\int_{-\infty}^{\eta}\rho v_{x}dz~dt\\
&=\frac{1}{2}\rho a^{2}\omega=\rho a^{2}k v_{g}\nonumber
\end{align}
If we start our wavemaker from rest and begin generating a wave we see that the mean wave surface must be elevated by $h=a^{2}k$.  

The energy flux contains three parts; the KE of reservoir particles, the PE of these particles and the $(PdA)v$ power needed to drive the existing particles out of the way.
\begin{align}
\mathit{f}_{K}&=\frac{\omega}{2\pi}\int_{0}^{2\pi/\omega}\int_{-\infty}^{\eta}\frac{1}{2}\rho (v_{x}^{2}+v_{z}^{2})v_{x}dz~dt
&=\frac{1}{4}\rho \omega^{3} a^{4}~~
&=\frac{1}{2}\rho g a^{2} (a k)^{2} v_{g}\\
\mathit{f}_{U}&=\frac{\omega}{2\pi}\int_{0}^{2\pi/\omega}\int_{-\infty}^{\eta}(\rho g z) v_{x}dz~dt
&=\frac{1}{8}\rho g \omega a^{4} k
&=\frac{1}{4}\rho g a^{2} (a k)^{2} v_{g}\\
{\text{P}}&=\frac{\omega}{2\pi}\int_{0}^{2\pi/\omega}\int_{-\infty}^{\eta}P v_{x}dz~dt
&=\frac{1}{4}\rho g \frac{\omega}{k} a^{2}~
&=\frac{1}{2}\rho g a^{2} v_{g}\\\nonumber
\end{align}
where the pressure $P=P_{w}+P_{g}=-\rho\frac{\partial}{\partial t}\Phi-\rho gz$ follows from Bernoulli's equation.  The vertically integrated KE is almost constant at all points of the wave.  In contrast, the PE density varies greatly yet we see that the source of this is not the PE of the injected particles.  The energy flux of the wave is primarily PE (see Sec.~\ref{Group}) but this PE is from work done on the particles already present by the pressure of the wavemaker.  

%IS IT A PROBLEM THAT THE POWER FLOWS ALL THE WAY ACROSS THE PACKET?  CHECK THE ANISOTROPIC GAS EXAMPLE AND SEE IF IT TRANSPORTS POWER.  

The momentum input has contributions from the injected particles and pressure on the edge of the wave being created
\begin{align}
\mathit{f}_{p}&=\frac{\omega}{2\pi}\int_{0}^{2\pi/\omega}\int_{-\infty}^{\eta}(\rho v_{x}) v_{x}dz~dt
&=\frac{1}{4}\rho g a^{2}
&=\frac{1}{2}\rho \omega a^{2} v_{g}~\label{Forces}\\
F&=\frac{\omega}{2\pi}\int_{0}^{2\pi/\omega}\int_{-\infty}^{\eta}P_{w} dz~dt
&=\frac{1}{4}\rho g a^{2}
& =\frac{1}{2} \rho \omega a^{2} v_{g}\\\nonumber
\end{align}
which gives twice the usual wavemaker result in Sec.\!~\ref{Wavemaker} without any integration constant ambiguity.  At the end of Sec.\!~\ref{Wavemaker} we will discuss forces on flapper-type wavemakers.

We can give an estimate of the effect of these unbalanced forces at the end of a packet by considering the case of a packet that is uniform in average height and drops to zero over one wavelength.  This means the force at the end of the packet is unbalanced and the last parcel of fluid in the packet is subjected to $F$.  In time $t=\frac{2\pi}{\omega}$ the end parcel advances at $v_{ph}$ but the packet only advances at $v_{g}=\frac{1}{2}v_{ph}$.  As in Sec.~\ref{Group}, we see that only half the elevated mass of the last half wavelength, $m=\frac{1}{2}(\rho a^{2}k \frac{1}{2}\lambda)$, should be advanced into this further half wavelength.  Let us compare this with the advance of this surface mass due to the pressure driven force $F$.
\begin{align*}
d\approx\frac{1}{2}\text{a}t^{2}
=\frac{1}{2}\frac{F}{m}t^{2}
=\frac{1}{4}\frac{g}{\pi}\frac{4\pi^{2}}{\omega^{2}}
=\frac{\lambda}{2}
\end{align*}
The consistency of this calculation suggests that the unbalanced force at the ends of a packet provides the extra mass transport (necessary by incompressibility and nonzero mass flux) to be consistent with packet spreading.  Therefore this force does not represent an extra momentum flux hidden in the packet motion.  If we simply grouped these terms together in a sweeping tensor equation we would not be able to see the distinction between \textit{momentum density flux} and stabilizing end \textit{forces}.  

%The units are the same but the meaning is completely different.  If we remove these end forces on a packet it simply begins to spread.  These are tailored oscillating forces as were our forces in the example of the anisotropic gas.  The work they provide in accelerating mass is recovered in the decelerating part of the motion.  It is not yet clear where such forces on a packet on the ocean could come from (if they are ever present).  
%
%This result makes us concerned that maybe the pressure driven power should not be included in the energy flux.  Our anisotropic gas example is no help here since it was noninteracting.  The energy of the wave created is $E=\text{P}\Delta t$ where $\Delta t$ is the time we have been running the wavemaker.  This  gives $E=\frac{1}{2}\rho g a^{2}(v_{g}\Delta t)=\mathcal{E}l$ which shows all the energy density comes from this power as expected.  

%All this shows that our usual intuition about waves and energy and our simple momentum density calculations are correct.  Let us now turn to angular momentum and see what the wavemaker imparts.  

There are kinetic and pressure driven contributions to the angular momentum input
\begin{align}
\mathit{f}_{L,KE}&=\frac{\omega}{2\pi}\int_{0}^{2\pi/\omega}\int_{-\infty}^{\eta}\rho (z v_{x})v_{x}dz~dt
&=-\frac{1}{8}\rho \frac{\omega^{2}}{k^{2}} a^{2}
&=-\frac{1}{4}\rho  a^{2} \frac{\omega}{k}v_{g}
=-\frac{1}{4}\rho g \frac{a^{2}}{\omega}v_{g}\\
\tau_{{P}}&=\frac{\omega}{2\pi}\int_{0}^{2\pi/\omega}\int_{-\infty}^{\eta}{P} z dz~dt
&=\frac{1}{8}\rho \omega^{2} a^{4} 
&=-\frac{1}{4}\rho g \frac{a^{2}}{\omega}(ak)^{2}v_{g}\\\nonumber
\end{align}
This gives $\mathcal{L}=-\frac{1}{4}\rho g \frac{a^{2}}{\omega}$
which is in agreement with our previous calculation.  

%Since we can turn off the wavemaker at will (when the surface height returns to zero), we know that no end transients with vertical momentum are introduced that would radically alter this result.  The pressure generated torque gives the torque required at the end of a packet to resist spreading.  This is evident since $\tau=(\frac{1}{2}a^{2}k\hat{z})\times F$ where we are applying the force $F$ to the c.m.\ of the $a^{2} k$ elevation of the packet.  

Before we move on to the kinematic implications of these conserved quantities let us consider the currently popular theory \cite{LH} on the ``radiation stress'' of waves and how much momentum is absorbed by the wavemaker that creates them.  This theory is built on some clever vector calculus manipulations of the N-S equations.  As such, it is somewhat opaque to physical intuition.  Our investigations in Sec.\!~\ref{Fluxes} suggests that there could be some hidden contributions to the flux that we have not included so we should investigate it carefully.  

\section{Wavemaker Theory and Radiation Stress}\label{Wavemaker}

One confounding notion in the subject of surface waves is the notion of ``radiation stress.''  This is described as the ``excess'' momentum flux associated with wave motion \cite{LH}.  It is often considered as a triumph of vector calculus methods in hydrodynamics to derive this as though it could not be predicted simply by other means.

Let us quickly review the analysis of the problem.  Reynolds transport theorem tells us how the momentum in a parcel changes in terms of the velocity changes in the incompressible fluid and surface terms:
\begin{align*}
\frac{d\mathcal{M}}{dt}=\rho\frac{d}{dt}\iiint_{\Omega}v~dV=\rho \iiint_{\Omega}\frac{\partial v}{\partial t}+\rho\iint_{\partial \Omega}v(v\cdot n) dS
\end{align*}
Assuming inviscid irrotational flow, the Euler equations imply:
\begin{align*}
\frac{d\mathcal{M}}{dt}=-\rho\iint_{\partial \Omega}\bigg(\frac{P}{\rho}n+v(v\cdot n) \bigg)dS
\end{align*}
where $n$ is the outward unit normal to the surface $\Omega$.  Now consider the volume $\Omega$ of a sea of progressive waves to be that bounded by the sea surface, bottom and a large vertical rectangle with two faces $S_{-}, S_{+}$ parallel to the wave fronts.
After arguing away the contribution to the x-momentum of the horizontal surfaces and those perpendicular to the wave fronts we obtain:
\begin{align*}
\frac{d\mathcal{M}_{x}}{dt}=\rho\iint_{\partial S_{\pm}}\bigg(\frac{P}{\rho}n+v(v\cdot n) \bigg)dS
\end{align*}
This gives the change in the x-component of momentum as a function of the pressure and velocity on the vertical surfaces that the waves cross.  Interpreting the momentum flux across each surface as
\begin{align*}
\frac{d\mathcal{M}_{x}}{dt}=-l\int_{-\infty}^{\eta(x)}\bigg(P+\rho v_{x}^{2} \bigg)dz\times\text{sgn}(n_{\pm})
\end{align*}
where $l$ is the length of the wave in the y-direction.  

Now we assume the surface $S_{-}$ is pushed far towards $x=-\infty$ where the wavemaker is located.  This lets us interpret the flux at $S_{+}$ as the momentum flux driven into the wavemaker.  Substitution and time averaging  gives 
\begin{align}
F=\frac{d\mathcal{M}_{x}}{dt}=-l\times \frac{1}{4}\rho g a^{2}
\label{LHforce}
\end{align}
as the force the wavemaker must absorb.  This is strange because the net mass drift which carries the momentum is opposite to this ``extra'' momentum of wave motion.  (There is a true mass flux that must be entering from the wavemaker and it is unclear why this is not occurring or if it is already figured into the calculation.)  We see that the wavemaker above does provide such a force.  %Of course, dimensionally such a result is practically forced on us and the coefficient out front has few possibilities since it comes from simple integrals.  Let us see if this is a kind of accident (as we will see in Sec.~\ref{Rigid}) or meaningful.  
%As we saw with the anisotropic gas and compressed fluid, the momentum flux does not need to be conserved to transport forces across them.  We also know to be wary of pressure contributions to the momentum.  In Sec.~\ref{Stress} we will see that there is no reason for us to expect the internal state of a medium to tell us about the momentum passing from one side of it to the other.  

One possible criticism is that we have taken the sum of the two terms that gave the momentum change in the volume bound by them and broken them up as the momentum that crosses each surface.  This is analogous to applying the fundamental theorem of calculus without remembering that the result is arbitrary up to a constant.  The vector calculus version is arbitrary up to a constant vector field (actually up to any vector field whose normal derivative integrates to zero on the surface).  To specify this value we must do an actual calculation of the momentum flux at one of the surfaces $S_{+}$ or $S_{-}$.  From the previous section we see that this has hidden complications and no special reason to think this extra constant field should be zero.  Furthermore, \textit{this arbitrary constant need not be the same from one configuration to another} so we can't try to extract it for a simple case and use it in one of greater interest.  Mistakes of this nature are usually hidden under the phrase ``let us interpret'' of some integrand rather directly evaluating the true physical quantity directly.  

As a second criticism, we notice that the sign of the momentum flux is entirely fixed by the the direction of the outwards normal and the positive definiteness of the arguements.  There is no way for this sort of result to ever allow flux in the opposite direction.  We have essentially hidden the $S_{-}$ surface at $-\infty$ or at the wavemaker.  %(where the dynamics are externally driven, not determined by the N-S equations).  
The time averaged result on $S_{-}$ gives exactly the opposite result to the one at $S_{+}$.  This, of course, indicates that the net momentum in our rectangle of sea is constant.  It is not clear that it says anything about the momentum passing through the system.  If we were to choose the surfaces to be separated by exactly an integer number of wavelengths, the two contributions are exactly equal and the value of the momentum of the region of sea between these surfaces is exactly zero at all times (regardless of time averaging).  The whole analysis began by investigating the change in momentum in a volume and manipulating the equations to get a surface contribution.  How are the equations that measure the momentum of a parcel with fixed momentum  supposed to tell us about the flux through the parcel?  %Even if we accepted that this is the momentum flux, we cannot just cut of the wave here and assume that this is the force on a solid surface there.  Such a truncation violates the conservation of mass at the boundary among other problems.  

If we were only interested in wave-wave interactions we might be able to \textit{define} the momentum of the waves by Eqn.\!~\ref{LHforce}.  These sorts of quasi-momentum definitions can be self consistent if we are not interested in exchanges of momentum with bodies outside the sea.  However, the wavemaker is exactly such a system.  Attempts have been made to quantify when real forces can be obtained from changes in quasimomentum \cite{Mc} but, even when these might work, we have the problem that surface waves at rigid objects get partially destroyed into shear flows, which exert their own forces and torques, not just those of reflected waves.  As we have seen in the angular momentum case, the clever approach can hold subtle confusions that the direct approach spares us.  
Based on this analysis, it is hard to see how this calculation contributes anything to our knowledge of the momentum flux of water waves.  More details on how to properly obtain the momentum flux and its nonadvective source terms from the equations of motion are in App.~\ref{Flux}.   
%Not having to worry about some hidden contribution means that we can be confident that a consideration of the the true momentum density is sufficient for a conservation law based study of the response of waves to wind forces. 

\section{Wave Tanks}\label{Wavetanks}

%WE SHOULD INCLUDE AN IDEAL FLAPPER TANK WITH A BACK WALL ELEVATION AND IRROTATIONAL FLOW AND A SOURCE AND SINK OF VORTICITY.  THE RETURN IRROTATIONAL FLOW JOINS WITH THE SOURCE FLOW.  ABSENT BREAKING, ALL VISCOSITY IS SOURCED AT WALLS.  
%
%Direction of flux depends on angle of shore.  Net L is clockwise.  No universality in L, p conversion to flow versus absorption by shore.  

Now that we have a clear understanding of the conservation laws and the stresses induced by wave motion (2D) let us apply them over the next few sections to some common wave problems.  
Flapper driven wave tank experiments have been the testbed for most experiments on Stokes drift.  The obvious problem is what happens to the supposed drift once the waves hit the other side of the tank.  Less obvious is where the mass flux for the wave is supposed to come from when the flapper initiates the wave.  Waves incident on a shore generate undertow, rip currents and a general elevation on the shore called wave setup that is a distinct contribution to the elevation caused by storm surge.  Let us briefly consider a simple case of wave setup before we move on to more typical tanks.   

The nature of wave setup has been investigated from the standpoint of radiation stress \cite{LH}.  We now have reason to question the validity of results derived from both perspectives though the role of torques during wave destruction certainly seems crucial.  Let us consider the problem from a very idealized perspective based on constraints and see what it can imply about the realistic situations. 

Consider the case of a simple flapper driven wave tank.  We know that the advancing waves to the right should carry mass and momentum but the wall prevents such a flow from persisting.  This means that waves are  destroyed and reflected, in some combination, with corresponding deposition of mass and momentum and dissipation of energy.  The conservation of mass requires that some backwards shear flow be established to cancel the drift.  This must be locally true so that we should get some ccw vorticity and mass return introduced near the surface as the flapper drives flux away at the surface.  
The details of such a process are turbulent and messy so let us modify our tank and its driving and dissipation processes to make the calculation easier.  

\begin{figure}[!ht]
   \centering
   \includegraphics[width=4in,trim=0mm 80mm 0mm 110mm,clip]{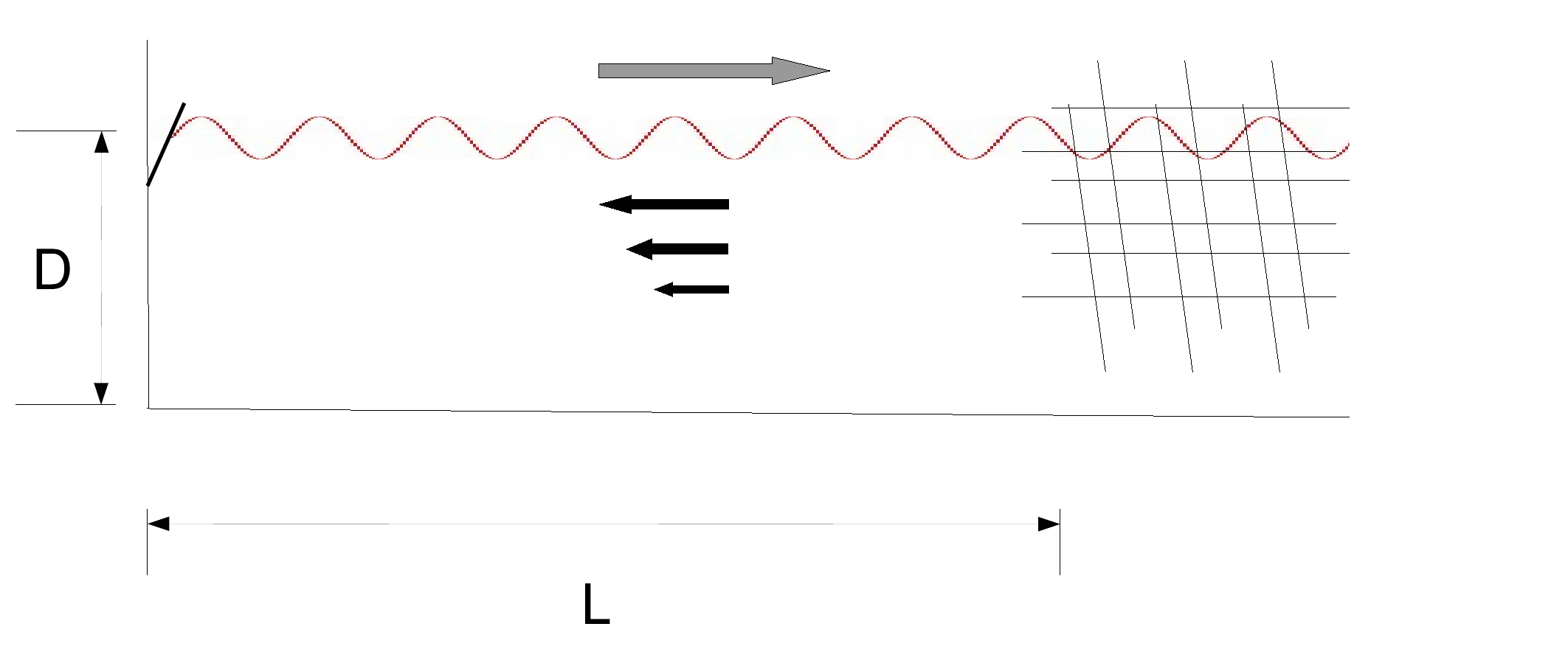} 
   \caption{A wave tank with waves generated at the left and moving right.  Waves are destroyed at the right grating and transfer maximal $p$ and $L$ to it consistent with the constraint of steady state mass flux.}
   \label{Wave-tank}
\end{figure}

Instead of a tank with a sloping beach or an abrupt wall on the opposite side to the wave flapper, consider a long grating of rods designed to destroy the waves with turbulence gradually over many wavelengths as in Fig~\ref{Wave-tank}.  This could be tailored to ensure virtually all of the wave goes into turbulent destruction instead of reflection or shear flows.  %We also can imagine measuring torques and forces on this grating to determine the angular and linear momentum imparted.  
%Wavetanks equilibrate on short time scales so we can measure the stationary behavior.

We assume that the Reynolds number of the free part of the tank is sufficiently small,  
and the turbulence in the grating region is localized so it doesn't introduce forces into the flow. From this we can make a simple model based on mass conservation to describe the flow profile and inclination of the water surface.  This is a kind of ``minimal'' configuration for wave-set up.  

The energy dissipated by the waves is large even though the mass transported by them is relatively small.  One can see  that the particular nature of the destruction could introduce large torques on regions of the fluid and minimize the heating losses of turbulence in favor of producing energetic flows.  This example seeks to maximize those losses and minimize the resulting forces.  The grating and the walls have an unlimited capacity to absorb linear and angular momentum and the fluid can absorb unlimited amounts of kinetic and potential energy as heat.  However, mass cannot be removed from the system like these other quantities.  There must be a backwards flow of mass to cancel the drift of the waves.  If the drift at the surface did not cancel we would need some extra recirculation.  

In the spirit of seeking the solution with the most energy possible dissipated at the grating, we search for conditions to give the simplest, least dissipative flow.  
The waves travel right and carry mass in the Stokes drift near the surface.  The mass build up creates laminar flow to the left that spreads out to minimizes dissipation (hence becomes a linear profile).  These transport rates must match.   The backwards laminar flow must be driven by a slope in the fluid surface (also linear).  The gravitational force on the sloping surface must be balanced by the viscous drag.  Therefore, we assume a tank of length and depth $L\gg D\gg k^{-1}$ that satisfies the following stationary conditions  
\begin{enumerate}
\item Flow at the bottom vanishes
\item Shear component of the flow is linear, $w=w_{0}+w_{1}z$.
\item The net flow (Stokes drift + flow) across any vertical section vanishes
\item $F_{\text{visc}}=-F_{\text{grav}}$
\end{enumerate}

The conservation condition leads to
\begin{align*}
w=-\frac{a^{2}\omega}{D}\left( 1+\frac{z}{D} \right)
\end{align*}
Matching the average power per volume supplied by gravity and dissipated by viscosity
\begin{align*}
\text{P}_{g}&=\text{P}_{\eta}\\
\rho g \left(\frac{\partial w}{\partial z}\right)\frac{D}{2}\sin(\theta)&=2\eta\left(\frac{\partial w}{\partial z}\right)^{2}
\end{align*}
where $\theta$ is the angle of inclination of the surface.  This gives
\begin{align}
\sin(\theta)=4\frac{\eta a^{2} \omega}{\rho g D^{3}}=8\left(\frac{w_{0}^{2}}{2 g D}\right)\frac{1}{Re}=4\frac{\text{a}}{g}\left(\frac{a}{D}\right)^{3}\frac{1}{Re}
\end{align}
where $Re=\rho w_{0} D/\eta$ is the Reynolds number of the flow, $w_{0}$ is the surface velocity of the shear flow, and $\text{a}=a\omega^{2}$ is the maximal acceleration of the wave motion.

A famous experiment by Bagnold \cite{Bagnold} used a tank 30~cm deep and sent waves 3~cm high and 1~m long towards a sloping ``beach'' to study the velocity profile of waves.  He found a $\sim$3~cm/s forwards flow at the bottom and a much slower backwards circulation at higher depths.  This gives a flow circulating, not surprisingly, with the same direction as the angular momentum of the waves.  Comparable waves  in our case would give a {backwards} surface flow of 2.2~cm/s and a slope of only $10^{-7}$!   This shows how strongly the remnants  torques that are not fully absorbed at the walls can affect the results.\footnote{$Re\ge 1000$ here so we are near the turbulent regime and our laminar approximation is not perfectly valid.  Additionally the deep water approximation should have some corrections as well.  The idealizations we have made would require a very long tank and a very smooth diffusion of flow out of the grating.  Given the very slight inclinations it produces, this is best considered as a pedagogical example.}  

\begin{figure}[!ht]
   \centering
   \includegraphics[width=4in,trim=0mm 180mm 0mm 150mm,clip]{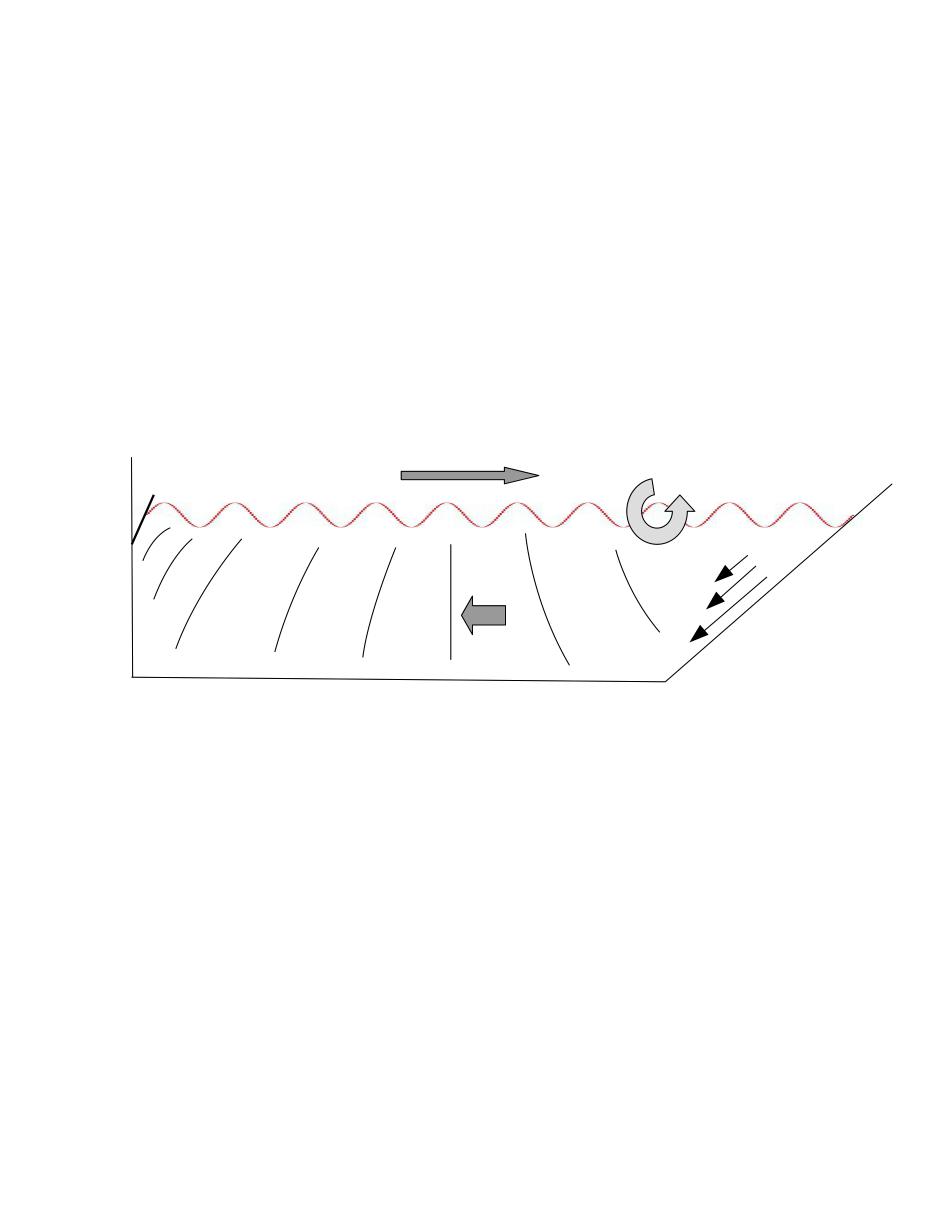} 
   \caption{A wave tank with flapper and slope for breaking of waves.  The direction of angular momentum of waves relative to surface is labelled along with the velocity potential of the irrotational back flow from the breaking region to the the flapper.  Part of the energy loss is into a vorticity source shear flow on the incline.  }
   \label{sources}
\end{figure}

The moral is that realistic wave set-up must be driven by the forces at the shore during wave destruction and an incomplete transfer of angular momentum rather than considerations based on linear momentum flux.  More realistic wave tanks modeled on ocean waves give an inclined wave breaking region as in fig.\!~\ref{sources}.  In this case there is generally a partial reflection of waves.  A the incline there is a transfer of linear and angular momentum to the container but the details of how much and how much vorticity in the flow is produced is very configuration dependent.  An irrotational backflow carries fluid back to the driver.  Details of the incline at the surface determine the direction and strength of the vorticity carrying shear flow that makes up the rotational component of the flow.  

%Suppose we wish to construct tank to reproduce the Stokes waves or predicted rotational waves.  
%The mass flux must be matched and instabilities and turbulence that drive surface flows eliminated.  This means we must include sources and sinks of fluid at the driver and the end of the tank.  Managing the turbulence introduced by jets on the driver could be quite challenging.  An alternate method would be to drive the wave with a conductive fluid and the MHD tools developed for ship propulsion.  These could impel the water in a more continuous fashion.  
%
%Another possibility is to drive the waves with surface air jets.  We can reshape the surface with jets that are vertically directed and translate them at a desired speed.  In Sec.~\ref{Driven} we will see that the wind driven waves in nature have constraints on $\mathit{f}/\tau$ and $\mathit{f}/\dot{\mathcal{E}}$.  This extra control of surface forces means we can be more free in how much of these quantities we impart to the system.  This control of the surface forces suggests we can drive waves in a more clever fashion that avoids instabilities and their associated shear flows and rapidly moving long waves that carry off angular momentum.  
%\begin{figure}[!ht]
%   \centering
%   \includegraphics[width=4in,trim=0mm 110mm 0mm 60mm,clip]{wavedriver} 
%   \caption{Vertical wind jets shape the surface and translate.}
%   \label{wavedriver}
%\end{figure}

It seems that there is no universality in the kinds of forces realistic wavemakers will exert on the wavemaker itself and the walls of the container.  However, we can use what we now know about wavemaker theory to place bounds on what energy, momentum and angular momentum must be imparted by a flapper or surface winds to drive the waves.  The surface instabilities and turbulence at the flapper will introduce additional contributions.   Assuming the driver can be engineered to be ``smart'' enough to minimize this, has a source reservoir of fluid and the far end can efficiently absorb the wave and excess mass flux, then  the force exerted by it is solely from the mass flux of the Stokes drift and the force of the harmonic pressure increase of the waves $F_{\text{ideal wavemaker}}=f_{p}+F_{{P}}=2f_{p}=\frac{1}{2}\rho g a^{2}$.  (This is, of course, in excess of the force it exerts to hold up the water to its equilibrium surface.)  In the case of a flapper driven device where the fluid mass must come rises upwards from a deep shear flow, this flow hits the wall then rises up vertically from below the flapper or hits the flapper directly.  This gives an extra flux $\lesssim f_{p}$ for the wavemaker to absorb.  In this case 
$F_{\text{flapper wavemaker}}\le 2f_{p}+F_{{P}}=\frac{3}{4}\rho g a^{2}$. 

%THERE IS NO UNIVERSALITY IN WAVE SETUP BY THE RESULTS OF RATCHET DAMPING.  FIND THE NET P, L RETURNED TO THE FLOW BY VORTICITY.  

\section{Conclusions}
Considering how ubiquitous the role of momentum flux is, the concept has some surprising subtlety involving hidden contributions, source terms and interesting boundary effects.  This should be a source of joy to those who relish the physical understanding of nature in such terms and a source of consternation to those who simply want to apply standard formalism and generate perturbation expansions or seek the shortest route to validate these.  Given the ubiquity that hydrodynamic stress has gained in calculations of wave set-up, acoustic streaming, and wave-shear interactions, it will come as a surprise to some that such a notion is often intrinsically inadequate for such purposes.  Others, who have always found something unsettling about such uses or just accepted it in quiet resignation, may take some comfort in an exposition that affirms this.  

Wave creation by wind is still a topic of some debate and the relation of observations to those of wavemakers in the lab is often poor.  Through an ``ideal'' wavemaker construction we have calculated that the force on the wavemaker is twice what the usual momentum flux calculation indicates and that realistic flapper wavemakers may give up to three times this value.  The usual wavemaker analysis is analyzed and the momentum flux it generates is seen to be a pseudomomentum and its utility in generating real forces for either ideal or realistic wavemakers seems very doubtful.  

Vorticity and its production is an essential feature of converting irrotational motion into streaming motion and other currents.  Vorticity and such flows generally move very slowly in comparison to waves.  The net equilibrium condition of waves on a shore or waves in a wavemaker will be a complicated function of the geometry and process of wave generation.  The usual radiative wave stress approach can be thought of as a hopeful attempt to generate some universal answers where none exists.  Such an approach is hardly reprehensible.  We should allow some liberty in initial approaches to problems but to continue to hold to such an approach long after there is reason to suspect it is neither working well nor consistent seems characteristic of some cultural deficiencies surrounding the subject.

\begin{appendix}
\section{The Density Flux}\label{Flux}
We have seen two examples, the fluid momentum flux tensor $\Pi_{ij}$ and the local wave momentum flux, where by manipulating the equations of motion we end up extracting a inequivalent results that both give global momentum conservation.  We can consider the density flux of of any quantity that is a function of $v$ like vorticity, $\omega=\nabla\times v$, and even nontensorial, nonconserved and nonanalytic expressions like $|v|$ or $v^{2}\hat{v}$.  We can define density flux matrices for such entities but they will, in general, not be tensors and may have singular contributions.   
%because of (1) neglected boundary conditions and (2) a carelessness in investigating how momentum is actually transported.  It is appealing to rely on very formal approaches.  Results can follow quickly and very little leg work is required investigating the conceptual meaning of results.  
%If the physics turn out not to be generally describable by the assumed mathematical structure we should not expect these methods to work.  Let us investigate the flux of quantities in an incompressible fluid with an eye to how well analyticity of the fluxes and tensoriality associated stress components hold in general.  
%Any local quantity that is purely a function of the velocity, i.e.\ $f(v)$, is advected because if we add a translational component to the velocity of the fluid $v(x)\rightarrow v(x)+u$ then the quantity has its motions altered by the constant velocity $u$.  The equations of motion will generally determine the evolution of such a quantity.  
The role of the pressure in an incompressible fluid is to create source and sink terms that transport portions of it in a nonlocal fashion.  This is all directly derivable from the equations of motion.  This is in contrast to classical field theories like EM where we can locally keep track of these quantities but there is no notion of advection.  In other words, it is sometimes ambiguous to say how fast the energy or momentum is moving (but not in the case of packets).  We have no way to tag parcels of these quantities.  

The case of a classical fluid is special in that there is a collection of massive particles that carry these quantities in their motion and forces.  We can keep track of the particles as though they were tagged.  There are two things we must keep in mind.  First, the nonlocal transport introduced by incompressibility or other elastic forces.  These convert kinetic motion into elastic energy which can create force gradients that are unlike the trackable transport of KE and momentum through discrete collisions in a gas.  Second, we must be wary of transport that is not tied to the collective average motions of the constituents that define the parcels and our macroscopic $v$.  As in the case of our anisotropic gas, this can mask hidden transport.  We have already argued that in the case of surface waves we don't need to worry about this but we can't guarantee it will be so generally.  

%Let us consider the case of 
%
%If we know that that the desired quantities in a flux density are advected and we know the eom it seems that we should be able to extract the momentum flux tensor from the two of these.  This is true and shown in the following procedure.

Consider a tensor (or matrix) valued density of interest $f=f(v)$, as in the case of momentum, that satisfies the following identity as a consequence of the incompressible N-S equations:
\begin{align*}
\frac{\partial f}{\partial t}=\nabla\cdot J+Q
\end{align*}
%\begin{align*}
%\frac{\partial f}{\partial t}=\nabla\cdot J+Q
%\end{align*}
where $J$ and $Q$ are functions of $v$.  
Since $f$ is locally defined in terms  the velocity field it is an advected quantity with flux density $T=f\otimes v$.  By substitution we obtain 
\begin{align*}
\frac{\partial f}{\partial t}&=-\nabla\cdot T+\bigg(\nabla\cdot(J+T)+Q\bigg)\\
&=-\nabla\cdot T+S
\end{align*}
%I THINK THIS SHOULD BE IN THE FORM OF A CONVECTIVE DERIVATIVE OPERATOR INSTEAD.  
%SHOULDN'T I HAVE A VDOT GRAD F TERMS IN HERE???
where the tensor $S$ represents the source and sink terms due to nonadvective transport of $f$ i.e.\ pressure and  external force transport.  Since $\nabla\cdot v=0$ we have $\nabla\cdot T=v\cdot\nabla f$ which gives
\begin{align}
\frac{\partial f}{\partial t}+v\cdot\nabla f =\frac{D f}{D t}=S
\end{align}

\section{Gerstner Waves}\label{Gerstner}
Here we briefly illustrate a case involving the role of vorticity in waves and conservation laws.  Gerstner waves \cite{Gerstner} are exact solutions of the N-S equations that give surface waves that contain vorticity.  Historically, they predated the Airy wave theory.  They have the novel feature that all the particles move in circles so there is no Stokes drift.  They are also the unique periodic wave solutions in which all particles remain on isobars \cite{Kalisch}.  

Given a body of water with surface elevation and velocity field it is not clear how one can generally separate ``wave'' from ``flow'' motion.  For periodic wave solutions we can investigate them as shortened packets and see if they propagate unchanged.  Waves tend to travel much faster than the underlying fluid motion.  The vorticity transport theorem tells us that vorticity is advected with the fluid so there seems to be no mechanism for the vorticity in a rotational wave to get carried along with a fast moving wave packet.  For such packet to persist, instead of separating into traveling irrotational waves and relatively slow localized surface shear, we need an extant surface shear for the packet to move through.  It can then draw on this as an uninterupted source of its voriticity.  

A small (right moving) Gersnter wave has the same dispersion relation, hence wave velocity, as a corresponding Airy wave but has no drift.  Consider a packet of such a wave of length $l$ in a flat sea that.  In the complementary region let there be a shear profile $v=-a^{2}\omega k e^{2 k z}$.  This is exactly the shear we need to cancel the Stokes drift for a corresponding Airy wave.  Unlike in the Airy wave case, no net packet elevation is required since no mass is transported.  It is unclear if the nonlinear pressure fields will conspire to preserve this or if some external constraints are required.  If there is some net elevation change on the order of $h=a^{2}k$, as with Airy waves, then such a configuration is not maintainable and the packet will break up into other components.  

%This lets us approximate the Gerstner wave as an Airy wave moving through a sea with surface drift that cancels its drift as it passes.  

%The shear flow induces a backwards transport of mass that is cancelled by the wave motion of the advancing wave.  This implies that there must be a moving surface elevation depression of the moving packet.  To match the mass flux we have $h=-a^{2}k$, exactly the opposite elevation change of an advancing Airy wave packet in a body of water otherwise at rest.  
\end{appendix}

\end{document}